\documentstyle[twocolumn,epsfig,aps]{revtex}
\draft

\begin{document} 
\twocolumn[\hsize\textwidth\columnwidth\hsize\csname
@twocolumnfalse\endcsname \title{Tilt Texture Domains on a Membrane
and Chirality induced Budding}
\author{Sarasij R. C.$^{1,2}$\cite{SAR} and Madan Rao$^{1,3}$\cite{MAD}}

\address{$^1$Raman Research Institute, C.V. Raman Avenue,
Sadashivanagar, Bangalore 560080,
India\\
$^2$Institute of Mathematical Sciences, Taramani, Chennai
600113, India\\
$^3$National Centre for Biological Sciences, UAS-GKVK Campus,
Bellary Road, Bangalore 560065, India
}

\date{\today}

\maketitle

\begin{abstract}
We study the equilibrium conformations of a lipid domain on a planar fluid membrane
where the domain is decorated by a vector field representing the tilt of the stiff
fatty acid chains of the lipid molecules, while the surrounding membrane is fluid and
structureless. The inclusion of chirality in the bulk of the domain induces a novel
{\it budding} of the membrane, which {\it preempts} the budding induced by a
decrease in interfacial tension.
\end{abstract}

 ] \vskip1.0in

The {\it budding} of membrane vesicles and subsequent {\it fission} is the only means of
traffic between compartmental boundaries within a living cell \cite{CELL}. Budding of
vesicles clearly requires membrane deformation --- coat proteins such as Clathrin and
COPs I-II have been postulated to play a crucial role in facilitating this deformation.
However the precise role of these coat proteins in budding events remains poorly
understood \cite{TRAFFIC}.

Membrane deformation leading to budding may also be triggered by changes in the
structure and composition of lipids. An oft-quoted mechanism, especially in the context
of artificial membranes, is the phase segregation in multi-component systems which leads
to budding due to a decrease in the interfacial energy\cite{BUD,SACKMANN} between the
lipid components. However the resulting bud has a size $R_{bud} > 2 \kappa/\lambda$,
where $\kappa$ is the bending modulus of the membrane and $\lambda$ the interfacial
tension, which for typical lipids is of order $5\mu$m, {\it much too large compared to
the budding events in the cell}. In this letter we propose a novel budding mechanism in
a multi-component lipid membrane, where one of the species is chiral and acquires a
tilt. This {\it chirality-induced budding} can result in a bud size as small as
$500$\AA.

Consider for simplicity a lipid bilayer membrane, where the upper layer is built of two
species, a majority component ({\it e.g.} DPPE) and minority component ({\it e.g.}
sphingolipid), while the lower layer consists of a uniform density of DPPE\cite{RAFTS}.
Let the minority species (sphingolipid) acquire a tilt with respect to the local normal
of the membrane. The tilt \cite{TILT} could be intrinsic (when the temperature $T <
T_m$, the main transition temperature of the chosen lipid) or acquired due to
association with a third component ({\it e.g.} cholesterol). We will work in the
temperature regime where there is at least a microphase separation of the two lipid
components over a length scale $R$ of the order of say $500$\AA. Note that most cellular
lipids are chiral and so, on acquiring a tilt, this molecular chirality will express
itself at length scales much larger than molecular dimensions. The chiral strength may
of course be enhanced by forming complexes with cholesterol \cite{MCCONELL}.

To start with, let us assume that the sphingolipid complex is spread, with a fixed,
uniform density, across a domain of area $A$, small enough so that over this scale the
membrane may be assumed flat.  As a result, our description entails a decoration of a
planar domain by a two-dimensional (2d) unit vector field $\bf{m}$, representing the
projection of the stiff lipid chains onto the plane of the membrane\cite{NOTE1}, while
the surrounding membrane is fluid and structureless.

Our low energy hamiltonian for
the membrane tilt domain is given, to quadratic order in the fields by
$E=E_B+E_L$\cite{SALANGER}, where the bulk energy
\begin{equation}
E_{B}  =  \int_{A}
\,k_{1} (
\nabla\cdot{\bf{m}})^{2}+k_{2}(\nabla\times{\bf{m}})^{2}
    + k_{c}(\nabla \cdot{\bf{m}})(\nabla\times
{\bf{m}}) ,
\label{eq:bulk}
\end{equation}
and the interfacial energy,
\begin{equation}
E_{L} = \oint dl\,\left(\sigma_{0}+\sigma_{1}(\bf{m} \cdot \bf{n})+
\sigma_{2}(\bf{m} \times \bf{n})\right)\, .
\label{eq:interface}
\end{equation}
The unit normal to the boundary $\bf{n}$ aims everywhere into the domain. Note that
$\nabla \times \bf{m}$ is a pseudoscalar in 2-dim and $k_{c}(\nabla \cdot \bf{m})(\nabla
\times \bf{m})$ is the bulk chiral contribution to the energy. The magnitude of $k_{c}$
is set by the density of cholesterol present in the sphingolipid complex. $\sigma_{0}$,
the isotropic line tension, is of the order of $10^{-8}$ dyne. The Frank constants
$k_{1}$ and $k_{2}$ are $\sim 10 k_{B} T_{room}$ while the interfacial energy of a
domain of radius $0.05 \mu m$ is of the same order\cite{DIPOLE}. This implies that
thermal fluctuations can be ignored at the temperatures of interest. We set $k_{1} = 1$
as our unit of energy.

The hamiltonian (\ref{eq:bulk}), (\ref{eq:interface}) has been studied in
\cite{SALANGER,PETTEY} in the context of textures on Sm-C$^{*}$ films and Langmuir
monolayers respectively. Apart from the context, there are three major points of
departure --- (i) our variational calculation and Monte Carlo (MC) simulations for the
$T=0$ phase diagram are `exact' giving results very different from \cite{PETTEY}, (ii)
we study the effect of strong anisotropic line tension and bulk chirality on the shape
and texture of the domain and (iii) we couple the tilt field to the local curvature of
the membrane and study the effect of chirality on membrane shape.
 
We rotate our coordinate axes and redefine the coefficient of anisotropic line tension
to obtain the simpler expression\cite{SALANGER,PETTEY} $E_{L} = \oint
dl\,(\sigma_{0}+\sigma(\bf{m} \cdot \bf{n}))$. The $T=0$ phase diagram is obtained by
minimising $E$ variationally subject to the constant area $A$ constraint. In all cases,
we have corroborated the results of our variational calculation by an MC simulation
using simulated annealing to avoid getting stuck in local minima. In this sense our
variational calculation is `exact'.

We first study the phase diagram when the isotropic line tension is extremely high
forcing the domain boundary to be circular.  To begin with, let us turn off bulk
chirality ($k_{c} = 0$) and set $k_{2} = k_{1}$. The $T=0$ phase diagram in the
$R-\sigma$ plane (where $R = \sqrt{A/\pi}$) has the following $4$ phases separated by
first-order boundaries (Fig.\ 1)\,:\,\, (1) {\underline {\it Boojum}} phase where the
centre of the circular domain is a distance $a$ away from the core of a virtual defect
of strength $N$. Placing the origin of polar coordinates $(r,\theta)$ at the core of the
boojum, we can describe the texture ${\bf m} \equiv (\cos \phi, \sin \phi)$ by the
equation
\begin{equation}
\phi = N \theta + c_{1} r \sin \theta + c_{2} r^{2} \sin 2\theta +
c_{3} r^{3} \sin 3\theta + \ldots\, .
\label{eq:texture}
\end{equation}
The energy $E$ is minimised with respect to the variational parameters $(N, a,
\{c_{n}\})$ for fixed $(R, \sigma)$. We find that the optimal value of $N$ is exactly 2,
independent of $a$. In drawing the phase diagram we parametrise the texture by $N$ and
$c_{1}$ alone; inclusion of $c_{2}$ and terms of higher order lowers the energy by 1\%
at most.  Fixing the area at say $R = 1$ (in units of a microscopic cutoff length
$r_c$), an increase in $\sigma$ leads to an increase in $c_1$ and pushes the core of the
virtual defect towards the domain centre, till at $a=1.7$ we pass into the annular
phase.

\begin{figure}
\centerline{\epsfig{figure=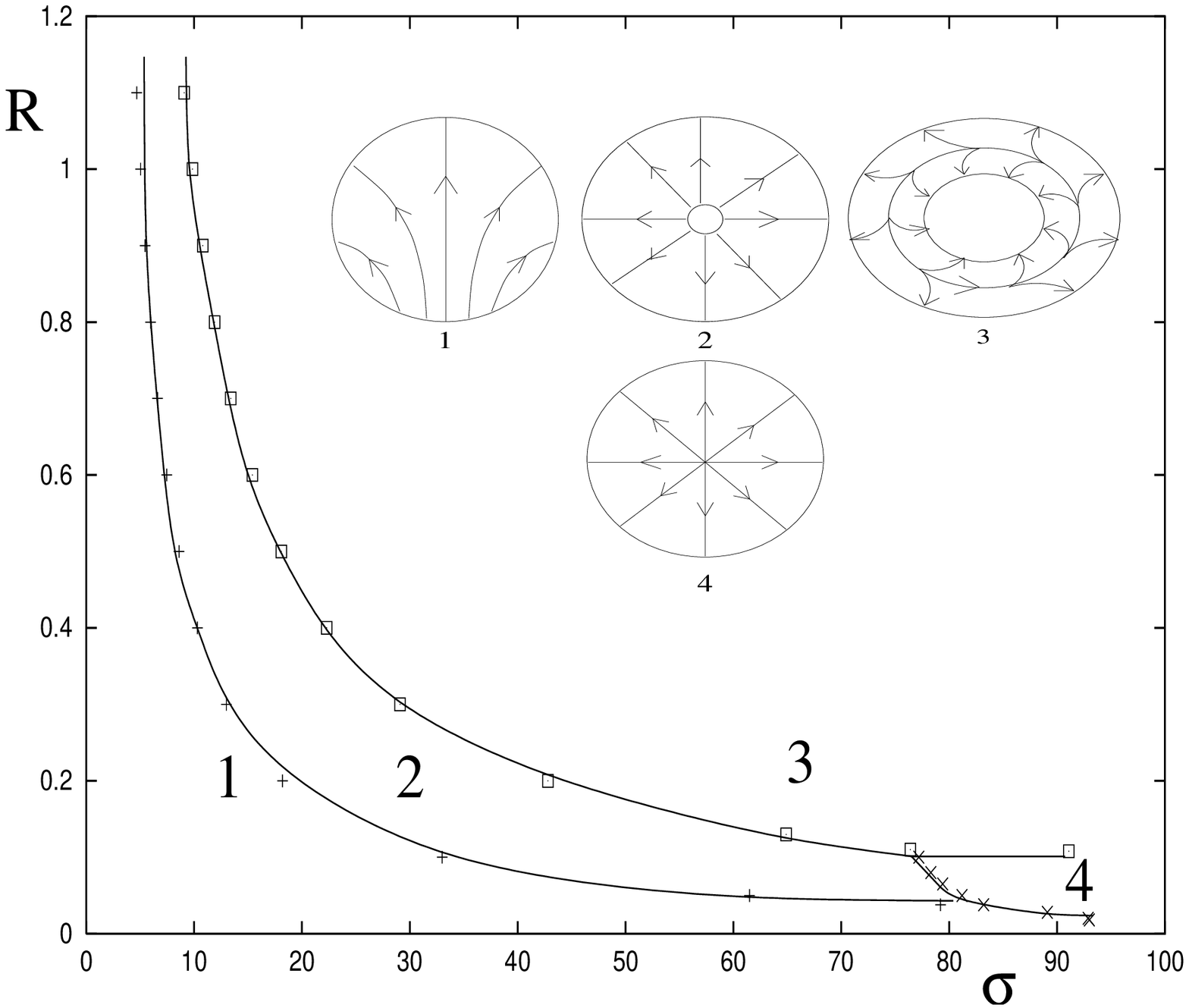,width=7.0cm,height=7.0cm}}
\end{figure}
FIG.1. Phases of the tilt texture domain with circular
periphery\,: (1) Boojum (2) Achiral Annular (3) Chiral Annular
(4) Hedgehog with
$\epsilon_{c} = 0$. Arrows indicate direction of {\bf m} field.
\\

\noindent
(2) {\underline {\it Achiral Annular}} phase with inner and outer radii $r_1$ and $r_2$
respectively. At the outer rim of the annulus $\bf{m}$ is directed radially outward
while at the inner boundary ${\bf m}$ is inclined at an angle $\alpha$ to the local
normal $\bf{n}$. With the origin of polar coordinates at the centre of the annulus, the
texture may be described by $\phi = \theta - \alpha (r_2-r)/(r_2-r_1)$, where $r_{1},
\alpha$ are variational parameters. The inner radius $r_1$ monotonically decreases as we
increase $\sigma$ such that $\bf{m}$ always points radially outward ($\alpha = 0$) at
every point of the domain. As one crosses the first order phase boundary (Fig.\ 1) we
encounter a new annular phase with a chiral texture.\,\, (3) {\underline {\it Chiral
Annular}} phase where $\alpha$ jumps to an angle greater than $\pi/2$ at both
boundaries. Consequently the bulk texture is chiral with the order parameter $C = \int
d^{2}\,{x}\,(\nabla \cdot \bf{m})(\nabla \times \bf{m}) \neq 0$ {\it even though there
is no bulk chiral interaction}. This chiral phase is doubly degenerate and can
spontaneously acquire either sign.\,\, (4) {\underline {\it Hedgehog}} phase with a
defect at the domain centre and a texture described by $\phi = \theta$. The energy
includes a core contribution $\epsilon_{c}$ of the defect of size $r_{c}$ .

The transitions between phases $1 \to 2$ and $2 \to 3$ are discontinuous (indicated by
the change in slope of the energy branches as a function of $\sigma$) and weaken as the
domain size $R$ shrinks.

As $\sigma_{0}$ is reduced, the boundary of the domain is no longer constrained to be
circular. We may then parametrise the boundary by smooth deformations of a circle,
$r(\theta)$ where the origin is at the centre of the circle of radius $r_0$ and $0 <
\theta < 2 \pi$ is the angle to the polar axis,
\begin{equation}
r(\theta) = r_0 \left(1 + \sum_{n=1}^{\infty} \alpha_{n} \cos (n\theta)
\right) \, .
\label{eq:boundary}
\end{equation}
The amplitudes $\{\alpha_n\}$ are additional parameters in our variational calculation.
We find that the texture throughout the $R-\sigma$ plane is a boojum with the domain
bulging out near the equator and flattening near the poles as $\sigma$ increases from
zero (henceforth we shall set $\sigma_0=1$, as a result our unit of length becomes of
order $0.1\,\mu$m). These shapes and textures win over the annular and hedgehog phases.

We now turn to a description of phases with nonzero bulk chirality $k_{c}$, which
without loss of generality we take to be positive. In the phase diagram, Fig.\ 2, we
have set $\sigma = 0$\,:\,\, (1) {\underline {\it Uniform}} phase, when $\vert
k_{c}\vert < 2$, where the domain is circular and the tilt field $\bf{m}$ is uniform.
(2) {\underline {\it Spiral Hedgehog}} phase where the domain is circular with a texture
such that $\bf{m}$ at every point is inclined at an angle $\alpha$ to the radially
outward ray emanating from the centre of the domain. The radial and the tangential
components of $\bf{m}$ in this spiral phase are $m_{r} = \cos \alpha$, $m_{\theta} =
\sin \alpha$, while the energy is given by
\begin{equation}
E = 2\pi (r_{c}+R) - (\frac{k_{c}}{2} \sin 2\alpha -
1) \log \frac{R}{r_{c}} + \epsilon_{c} \, .
\label{eq:energy1}
\end{equation}
The optimum value of $\alpha$ is $\pi /4$ when $\vert k_{c}\vert > 2$. This phase
clearly gives a nonzero value for the chiral order parameter $C$.  (3) {\underline {\it
Tweed}} phase domain is again circular though the texture loses the circular symmetry of
the previous phase. Our ansatz for the texture is motivated by the results of an MC
simulation of $3055$ particles carrying O(2) spins on a triangular lattice with a
hamiltonian obtained by discretising $E$, for $\sigma = 0$, $k_{c} = 25.5$. We have
performed a simulated annealing from $(k_BT)^{-1} = 0.1 \to 300$ starting from a variety
of initial conditions\cite{CUTOFF} to obtain textures as in Fig.\ 2 (lower inset).

\begin{figure}
\centerline{\epsfig{figure=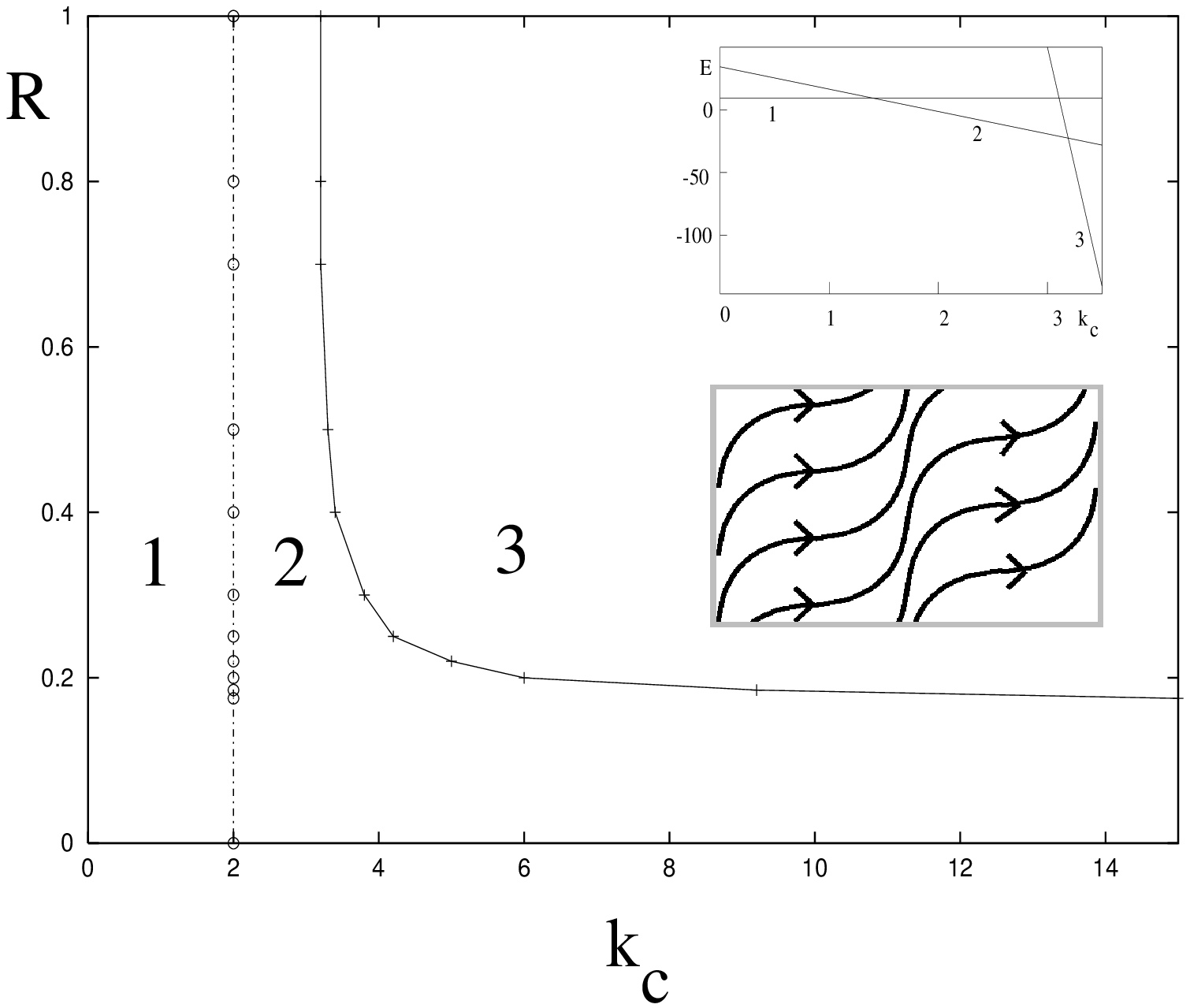,width=7.0cm,height=7.0cm}}
\end{figure}
FIG.2. Phases of the tilt texture domain with bulk chirality and $\sigma =
0$\,: (1) Uniform (2) Spiral Hedgehog ($\epsilon_{c}=0$) (3) Tweed. 
Top inset shows variation
of the energy branches as a function of 
$k_{c}$ at $R = 1$ indicating discontinuous transitions. Notice that the
energy
decreases rapidly in phase (3) as $k_c$ increases.
Lower inset shows the texture within a unit cell\,; this pattern repeats
periodically to form the Tweed phase (3).
\\

\noindent
A variational ansatz may be easily written down for such a texture: place the origin of
coordinates on the boundary of the domain and parametrise the texture by a length, $l$,
such that $m_{x} = \vert \cos x/l \vert$ and $m_{y} = -\vert \sin x/l \vert$. This
texture confers a net chiral strength to the domain ---
$(\nabla\cdot\bf{m})(\nabla\times\bf{m})$ has the same sign everywhere, though the sign
of individual terms, $\nabla\cdot\bf{m}$ and $\nabla\times\bf{m}$, varies from one
stripe to the next. The energy of the tweed domain is
\begin{equation}
E = 2\pi R -
\frac{k_{c}}{2l^2} \int_{0}^{2 R} dx \sqrt{2xR-x^2}\,
\vert \sin \frac{2x}{l}\vert + \frac{1}{l^2}\, .
\label{eq:energy2}
\end{equation}
The variational parameter $l$ approaches zero to minimise the energy, corresponding to
infinitely many tweeds. Higher order derivative terms in the hamiltonian would however
cutoff this monotonic decrease at some scale $l^{*}$\cite{CUTOFF}.

On turning on the anisotropic line tension $\sigma$, the boundary of the domain deviates
from a circle giving rise to a variety of new phases. We will make a detailed study of
the phase diagram in a later paper.  Let us move on, for the moment, and allow the
membrane containing the domain to be flexible, thus promoting a coupling between the
tilt field ${\bf m}$ and the curvature tensor $K_{ij}$\cite{HELPRO}. At the same time we
have to treat the membrane as a bilayer: we thus define the curvature, tilt and lipid
density fields on both the upper and lower leaves of the bilayer and project these
variables onto the neutral surface of the membrane\cite{SHILL}.  In covariant form, the
bulk energy to lowest order in derivatives, is given by \cite{NELSON},
\begin{eqnarray}
E_{B}  &  =  &  \int_{A} \left[ k_{1}({\bf{div}}\,{\bf{m}})^{2}+
            k_{2}({\bf{curl}}\,{\bf{m}})^{2}+
k_{c}({\bf{div}}\,{\bf{m}})({\bf{curl}}\,{\bf{m}})\right.  \nonumber \\
       &     &
\left. + c_{0} K_{i}^{i} + \kappa\, (K_{i}^{i})^{2}+
\beta \,m^{i}m^{j}K^{i}_{j} +
c_{0}^{*}\gamma_{ij}m^{k}m^{i}K^{j}_{k}\right]\, ,
\label{eq:covariant}
\end{eqnarray}
where the divergence and curl are written in terms of the covariant derivative $D_{i}$
as, ${\bf{div}}\,{\bf{m}} = D_{i}\,m^{i}$ and ${\bf{curl}}\,{\bf{m}} = \gamma_{i}^{j}
D_{j}\,m^{i}$. The appearance of $\gamma_{ij}$, the completely antisymmetric tensor on a
2d curved manifold \cite{HELPRO}, indicates that the coupling of the texture to the
membrane curvature also depends on the chiral strength of the sphingolipid complex
\cite{MCCONELL} in the domain. To highlight the effect of chirality, we set
$c_{0}=\beta=0$ and only include the isotropic line tension in the interfacial energy
$E_{L}$\cite{NOTE2}.

We will see that the effect of increasing $c_{0}^{*}$ for fixed size $R$ is to induce
the domain to leave the plane and form a spherical bud of radius $R/2$ connected to the
plane by an infinitesimal neck !  More significantly, at fixed but large enough
$c_{0}^{*}$ this {\it chirality induced budding occurs at a much smaller size $R$ than
the phase separation induced budding caused by a decrease in the interfacial tension
between the two lipid components}\cite{BUD}.

For instance if we fix $k_c$ at phase (2) of Fig.\ (2), we find that the membrane
remains {\it planar} and the texture a spiral hedgehog for small $c_{0}^{*}$ (Fig.\ 3).
A further increase in $c_{0}^{*}$ leads to a discontinuous budding transition where the
texture decorates a spherical bud with a vanishingly thin neck.  Both the chiral terms,
proportional to $k_c$ and $c_{0}^{*}$, gain in energy by this budding, even at the cost
of producing a defect at the pole. This may be seen by setting the latitudinal and
longitudinal components of ${\bf{m}}$ as $m_{\theta} = 1/\sqrt{2}$ and $m_{\phi} =
1/\sqrt{2}$, a choice which smoothly goes over to the spiral hedgehog phase in the limit
of the planar membrane. The membrane shape is parametrised as a sphere of radius $r$
attached to the plane by a neck of radius $r_0$. The energy of this conformation is
$E_{L} = 2 \pi r_{0}$ and
\begin{equation}
E_{B} = \pi \kappa \frac{R^2}{r^2}
- \pi c_{0}^{*} \frac{R^{2}}{r}
- \pi (k_{c} - 2) \int_{\theta_{c}}^{\pi - \theta_{0}} d\, \theta
 \frac{\cos^{2} \theta}{\sin \theta} 
\end{equation}
in addition to a defect core energy $\epsilon_{c}$, where $\theta_{c} = r_{c}/{r}$ and
$r_{0} = r \sin \theta_{0}$. We have ignored the small curvature energy cost at the
junction where the spherical bud meets the plane. Variation suggests that the energy is
minimised when $r_0\to0$ and $r=R/2$ (Fig.\ 3).

At larger values of $c_{0}^{*}$, the spherical bud gives way discontinuously to a {\it
cylindrical tubule} (Fig.\ 3). We see that the energy is minimized when the domain wraps
around a cylinder of radius $r$ and length $L$ attached to the plane at one end and
capped by a hemisphere with a polar defect core at the other (Fig.\ 3).  The ${\bf{m}}$
field on the spherical cap is as described earlier; the axial and tangential components
of ${\bf{m}}$ on the cylinder are $m_{z} = 1/\sqrt{2}$ and $m_{\phi} = 1/\sqrt{2}$. This
choice corresponds to a texture where the lines of ${\bf{m}}$ wrap around the cylinder
as a helix of pitch proportional to $r$ \cite{SELINGER}.

\begin{figure}
\centerline{\epsfig{figure=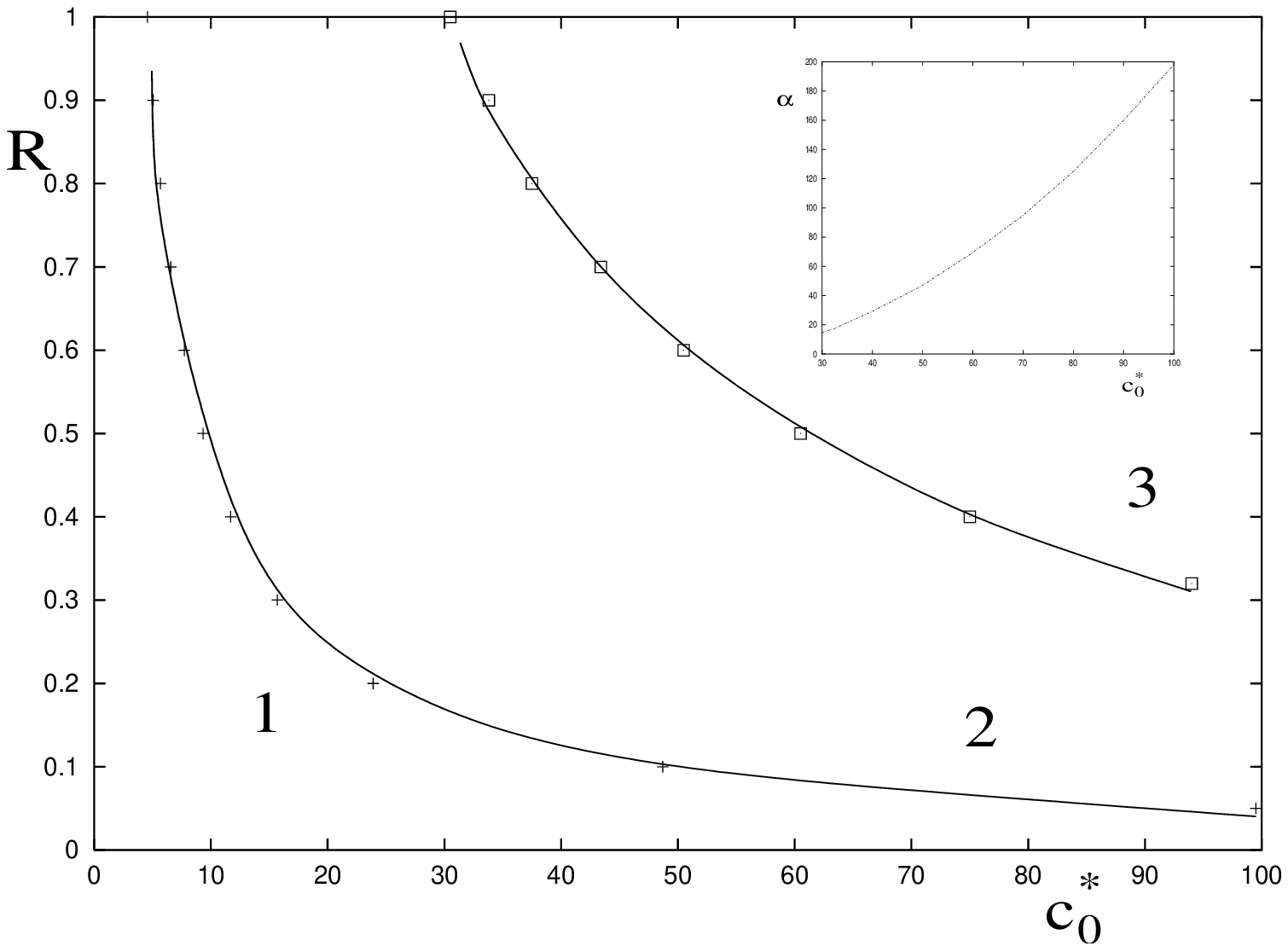,width=7.0cm,height=7.0cm}}
\end{figure}
FIG.3. Phases of the tilt domain coupled to membrane shape with $k_{c} = 2.5$,
$\sigma = 0$ and $\kappa = 5$\,: 
(1) Planar (2) Spherical Bud (3) Cylindrical Tubule.
With these parameters, domain induced budding due to a gain in line
tension\cite{BUD}
occurs at $R_c= 10$. Inset shows the variation of the ratio $\alpha = L/r$ with
$c_{0}^{*}$ for a domain of size $R = 1$. \\

\noindent
Note that ${\bf m}$ varies smoothly across the junction of the cylinder and the
hemisphere, and extrapolates smoothly to the spiral hedgehog texture in the limit of the
planar membrane. The energy of this conformation is $E_{L} = 2 \pi r$ and
\begin{eqnarray}
E_{B} &  =  &  2 \pi (\kappa - c_{0}^{*} r)
          + \pi \left(\frac{\kappa}{2} - c_{0}^{*} r \right)
            \left[\frac{R^{2}}{2r^2} - 1 \right] \nonumber \\
      &     & - \pi (k_{c} - 2) \int_{\theta_{c}}^{\pi - \theta_{0}} d
\theta \frac{\cos^{2} \theta}{\sin \theta} 
\end{eqnarray}
in addition to a core energy $\epsilon_{c}$ (we have again ignored the curvature
contribution coming from the neck). The optimal $L/r$ obtained variationally, increases
rapidly with $c_{0}^{*}$ (Fig.\ 3 (inset)).

As seen from the inset of Fig.\ 2, the energy plummets as $k_c$ increases. This implies
that if $k_c$ was chosen such that phase (3) of Fig.\ 2 were the equilibrium texture at
$c_{0}^{*}=0$, then the domain would gain significantly in bulk energy by budding, first
into a spherical bud and then into a tubule. The range of $c_{0}^{*}$ for which the
spherical bud obtains would however be reduced.

The variational calculation for the $T=0$ energy-minimising shapes of a flexible, fluid
membrane having a tilt domain is of course not exact. A more general variational
calculation will at most shift the phase boundaries slightly --- in any case the
prediction of a {\it chirality-induced budding} at small $R$ is robust and should be
testable in experiments on artificial membranes consisting of carefully chosen lipid
components. More importantly, we believe that this may have some relevance in the
budding of cell membranes which are not assisted by coat-proteins, as in the endocytosis
of raft components \cite{IL2R}.

We thank Yashodhan Hatwalne for several discussions and Sriram Ramaswamy for a critical
reading of the manuscript. This work would not have been possible without the insights,
criticisms and encouragement of Satyajit Mayor whose experiments initiated this line of
investigation. MR thanks DST, India for a Swarnajayanthi grant.


\begin{references}
\bibitem{NOTE1}
Since the membrane is flat, the bilayer nature plays no
significant role.
\bibitem{DIPOLE}
The dipole-dipole interaction energy per lipid is about $10^{-15}$ erg, two orders
smaller than the Frank energy (\ref{eq:bulk}) and hence may be ignored.
\bibitem{NOTE2}
Inclusion of $c_0$ and $\beta$ would merely favour the budding
transition owing to the intrinsic asymmetry of the bilayer.
\end{references}
\end{document}